\documentclass[onecolumn,showpacs]{revtex4}
\usepackage{amssymb}
\usepackage{amsmath}
\usepackage[dvips]{graphicx}

\begin{document}

\title{Molecular dynamics simulations of oxide memristors: thermal effects}
\author{S. E. Savel'ev$^{1}$, A. S. Alexandrov$^{1,2}$, A. M. Bratkovsky$%
^{2} $, and R. Stanley Williams$^{2}$}
\affiliation{$^1$Department of Physics, Loughborough University, Loughborough LE11 3TU,
United Kingdom\\
$^2$Hewlett-Packard Laboratories, 1501 Page Mill Road, Palo Alto, California
94304}

\begin{abstract}
We have extended our recent molecular-dynamic simulations of memristors to
include the effect of thermal inhomogeneities on mobile ionic species
appearing during operation of the device. Simulations show a competition
between an attractive short-ranged interaction between oxygen vacancies and
an enhanced local temperature in creating/destroying the conducting oxygen
channels. Such a competition would strongly affect the performance of the
memristive devices.
\end{abstract}

\pacs{71.38.-k, 74.40.+k, 72.15.Jf, 74.72.-h, 74.25.Fy}
\maketitle

There are many challenges in understanding and controlling the coupled
electronic and ionic kinetic phenomena dominating the behavior of oxide
switching devices like Pt/TiO$_{2}$/Pt, which is an exemplar memristor
(resistor with memory)\cite{stan}, a fourth passive circuit element
originally postulated by Leon Chua in 1971\cite{stan0}. It has been
demonstrated unambiguously that bipolar switching involves changes to the
electronic barrier at the Pt/TiO$_{2}$ interface due to the drift of
positively charged oxygen vacancies under an applied electric field\cite%
{stan}. Various direct measurements revealed formation of localized
conducting channels in TiO$_{2}$: pressure modulated conductance microscopy
\cite{joshG09,miaoG09}, conducting atomic force microscopy (AFM) \cite%
{ruth09},\ scanning transmission x-ray microscopy \cite{strachan10}, and
in-situ transmission electron microscopy \cite{kwon10}. On the basis of
these measurements, it became quite clear that the vacancy drift towards the
interface creates conducting channels that shunt, or short-circuit, the
electronic barrier and switch the device ON\cite{stan}. The drift of
vacancies away from the interface breaks such channels, recovering the
electronic barrier to switch the junction OFF.

The microscopic understanding of the atomic-scale mechanism and
identification of the material changes within an insulating barrier appear
to be invaluable for controlling and improving the memristor performance.
The number of oxygen vacancies within the volume 10$\times $10$\times $2 nm$%
^{3}$ of perspective nano-memristors could be as small as a thousand, so
that the conventional statistical (diffusion) approach for dealing with
many-particle systems might fail. However, the dynamic phase transitions as
well as the competition between thermal fluctuations and particle-particle
interactions in stochastic transport can be investigated with the use of
Molecular Dynamics (MD) simulations of the Langevin equations describing the
thermal diffusion and drift of individual interacting particles \cite{sav}.

Recently, we have proposed a model for the kinetic behavior of oxide
memristors and simulated its toy analog using the MD \cite{saaw}. Our MD
simulations revealed a significant departure of the vacancy distributions
across the device from that expected within a standard drift-diffusion
approximation. Another important step would be to include thermal effects
via local changes in the diffusion coefficient. Indeed, the channel
formation in systems like TiO$_{2}$, NiO, VO$_{2}$ is certainly accompanied
by local heating that is witnessed by the local emergence of
high-temperature phases and is observed by thermal microscopy. For instance,
fresh TiO$_{2}$ samples have an amorphous layer of titanium dioxide. To make
the device switch in a repeatable manner, a \emph{forming step} of rather
large voltage pulse is usually required to create a vacancy-rich region
(forming is not required if e.g. a vacancy-rich layer is provided
intentionally by fabrication). During this process, the TiO$_{2}$ anatase
phase forms near the conducting channel, which nominally requires
temperatures over 350$^\circ$C \cite{strachan10}. Studying local thermal
effects during switching in TiO$_{2}$ provides strong evidence for local
heating \cite{borg09}. Unambiguous evidence for local heating accompanying
conducting channel formation comes from third harmonic generation \cite%
{noh3w08}; low resistance states (LRS) in NiO showing \emph{unipolar}
switching were strongly nonlinear with variations in the resistance as large
as 60\%, which was most likely caused by the Joule heating of conducting
channels inside the films.
\begin{figure}[tbp]
\begin{center}
\includegraphics[angle=-0,width=0.85\textwidth]{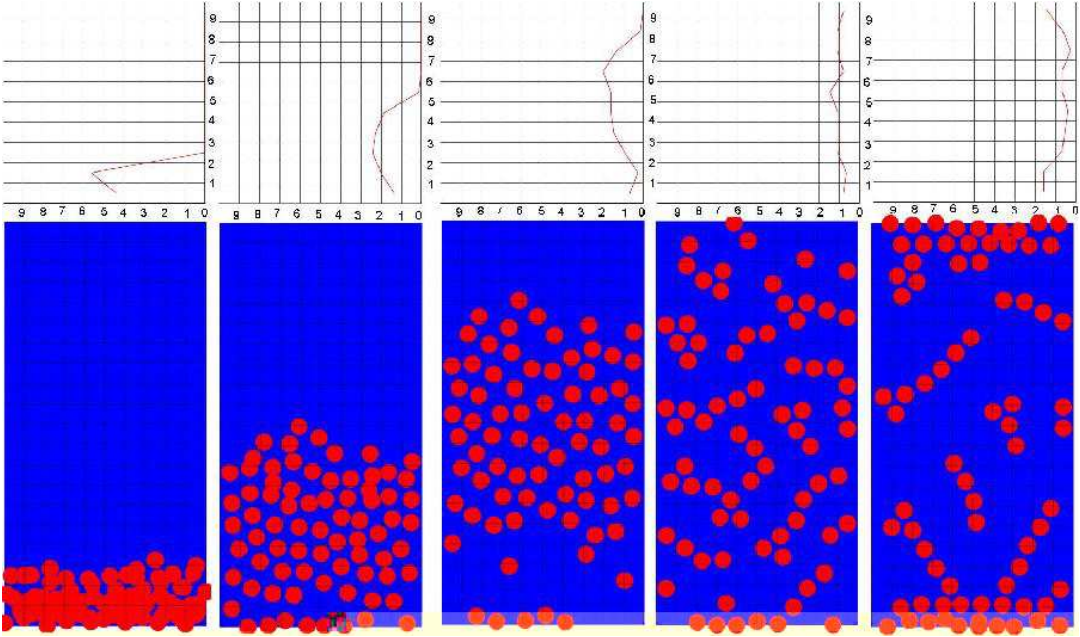}
\end{center}
\caption{(Color online) Simulation of dynamics of vacancies during
and after the external voltage pulse is applied. Parameters are the
same as in Fig.~1
in Ref.~\protect\cite{saaw}. Top row shows the distribution of vacancies $%
n(y)$ across the sample (with $y-$axes aligned with those in the
bottom row.) One can observe how the vacancies redistribute at
various simulation times ($t=0$, 0.5, 0.9, 1.5, 5 left to right, the
time measured in pulse duration). The final configuration (last
column, $t=5$) is a practically stable distribution with a larger
density of vacancies near the top and the bottom electrodes due to
the attractive interaction with the image charges. These patterns
are slightly different from Fig.~1 in Ref. \protect\cite{saaw} since
we have many metastable states that may trap the vacancies. }
\label{fig:1puls}
\end{figure}

Here, we extend our MD of oxygen vacancies interacting via realistic
potentials and driven by an external bias voltage by taking into account
temperature gradients in thin films of oxide memristors. Based on the
observations of oxygen vacancy migration and clustering in bulk \cite{miy}
and nanoscale \cite{kwon10,strachan10} samples of TiO$_{2}$ induced by an
electric-field, we have modeled a memristor \cite{saaw} as shown in Fig.~\ref%
{fig:1puls}. In our model, there is a reduced rutile thin layer TiO$_{2-x}$
near one of the metallic electrodes stabilized by the Coulomb mirror
potential. Vacancies from this layer can drift toward the opposite electrode
in an electric field. Vacancies interact with each other via the pairwise
potential $W$, and with external fields corresponding to the time-dependent
deterministic force $F$. The environment exerts an independent Gaussian
random force, $\vec{\xi}$ on each particle with zero mean and intensity
controlled by the temperature. In the overdamped regime (where inertia is
negligible compared to the viscous damping), the Langevin equation
describing the drift-diffusion of the i-th particle is \cite{sav}:
\begin{equation}
\eta {\frac{dx_{i}^{\alpha }}{{dt}}}=F_{i}^{\alpha }(\boldsymbol{x}%
_{i},t)-\sum_{j\neq i}{\frac{\partial W(\boldsymbol{x}_{i}\mathbf{-}%
\boldsymbol{x}_{j})}{{\partial }x{_{i}^{\alpha }}}}+\sqrt{2k_{B}T\eta }%
\gamma ^{\alpha }\xi _{i}^{\alpha }(t).  \label{lan}
\end{equation}%
where $x_{i}^{\alpha }$ is the $\alpha $-coordinate of the $i$th vacancy, $%
\eta $ is the friction coefficient, $F_{i}^{\alpha }$ is the $\alpha $%
-component of electric pulse force, $k_{B}$ is the Boltzmann constant, $%
\gamma $ is related to diffusion anisotropy, and $\xi ^{i}$ is a random
force.
\begin{figure}[tbp]
\begin{center}
\includegraphics[angle=-00,width=0.75\textwidth]{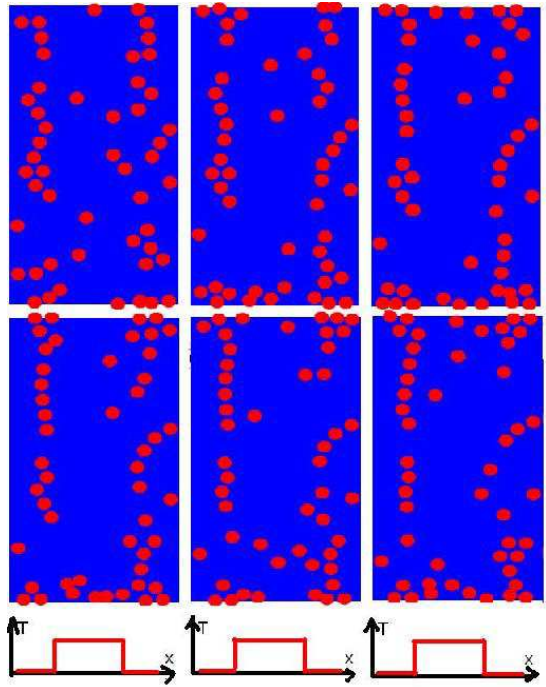}
\end{center}
\caption{(Color online). Vacancy distribution formed long time after
the
voltage pulse has been applied (top panel: $t=$5, 5.1, 5.5; bottom panel: $%
t= $5.7, 6, 6.5) where the time is measured in the pulse duration
interval. All parameters are the same as in
Fig.~\protect\ref{fig:1puls} apart from the anisotropy coefficient
$\protect\gamma =1/3$. Temperature varies across the sample with the
amplitude $T_{0}$ equal to the depth of potential minima
corresponding to the force (\protect\ref{force}). This temperature
is enough to break up the vacancy chains. } \label{fig:2Tx}
\end{figure}

The O$^{2-}$ vacancy-vacancy interaction potential, which is crucial for
their clustering and phase transformations, can be modeled as
\begin{equation}
W(R)=Ae^{-R/a}-\frac{B}{R^{6}}+\frac{e^{2}}{\pi \epsilon _{0}\epsilon R},
\end{equation}%
where the short-range repulsive and attractive parts are represented with
the parameters $A=22764.0$ eV, $a=0.01490$ nm, and $B=27.88\times 10^{-6}$
eV nm \cite{potential}, and the long-range Coulomb repulsion is given by the
last term. More generally, the interaction parameters $A,B,a$ vary from one
oxide to another, and the dielectric constant $\epsilon $ may also vary from
sample to sample.

In contrast with strongly inhomogeneous (internal) electric fields, we have
assumed in Ref.\cite{saaw} that the diffusion coefficient is homogeneous in
the simulated region, which corresponds to a homogeneous temperature
distributions. Here, we extend our model by taking into account local
overheating in the sample, which appears to have a drastic effect on the
vacancy distribution pattern Indeed, the diffusion coefficient exponentially
depends on temperature \cite{rad}, so that the variations of the local
temperature during switching could significantly affect the formation and
disintegration of conducting channels.

We have performed the MD simulations of a toy memristor with relatively
small number of vacancies, $N=75$ (Fig.~\ref{fig:1puls}). Initially, we have
placed all the vacancies randomly near the bottom of a sample (see Fig.~\ref%
{fig:1puls} for the initial distribution) and then let these vacancies
evolve according to Eq.~(\ref{lan}) inside a rectangular box that mimics the
actual device. We use the 2D simulation area with the aspect ratio $%
L_{y}/L_{x}=2$ and periodic boundary conditions (BC) along the $x-$%
direction. Note that the periodic BC allow us to simulate an infinite area
sample using a rather small number of particles. To use periodic BC, we
include periodic images of the vacancies with respect to vertical boundaries
of the simulation box. We also incorporate opposite polarity charges by
adding mirror images of vacancies with respect to the top and the bottom of
the sample.

Since we simulate a limited number of vacancies, one can refer to each
particle in our simulations as a cluster of vacancies, where a
cluster-cluster interaction may be described by the combination of the
Lennard-Jones and Coulomb potentials acting with the force
\begin{equation}
F(r)=\frac{1}{r}\left\{ 12E_{LJ}\left[ \left( \frac{r_{min}}{r}\right)
^{12}-\left( \frac{r_{min}}{r}\right) ^{6}\right] +E_{c}\frac{r_{min}}{r}%
\right\} ,  \label{force}
\end{equation}%
where the relative strength of the Coulomb potential is given by $%
E_{c}/E_{LJ}=2$. This results in the position of the potential maxima $%
r_{max}\approx 2r_{min}$ and the height of the potential barrier on the
order of the depth of the potential well. The pulse strength is taken to be
about ten times stronger than the maximum attractive force between vacancy
clusters.

We present the results of simulation for the device with \emph{homogeneous}
temperature distribution \cite{saaw} in Fig.~\ref{fig:1puls}. Different
columns refer to different moments in time and the top row shows the
corresponding local densities. The first column shows the initial
distribution of vacancies before simulations: all of them are located at the
bottom of the sample. The second and third columns show distribution in the
middle and at the end of the pulse. One can see that vacancies are pushed by
the pulse to the center of the sample: the density distribution is spread
out and its maximum moves towards the top of the sample with a constant
velocity. Moreover, even after switching the pulse off the vacancy
distribution front keeps moving towards the top of the sample while the
vacancies gradually form filamentary clusters (fourth column). Note that,
obviously, such filamentary clusters are impossible to reveal when modeling
vacancy dynamics by using just vacancy density $n(y)$. Indeed, one cannot
see anything interesting from the average density profile (top panel of the
fourth column). Finally, vacancies form the percolation paths that are
stable during all simulations while the average density again shows nothing
interesting apart from a trivial fact that $n(y)$ is slightly larger at the
top and the bottom of the sample due to the vacancies being attracted to
their image charges.

\begin{figure}[tbp]
\begin{center}
\includegraphics[angle=0,width=0.35\textwidth]{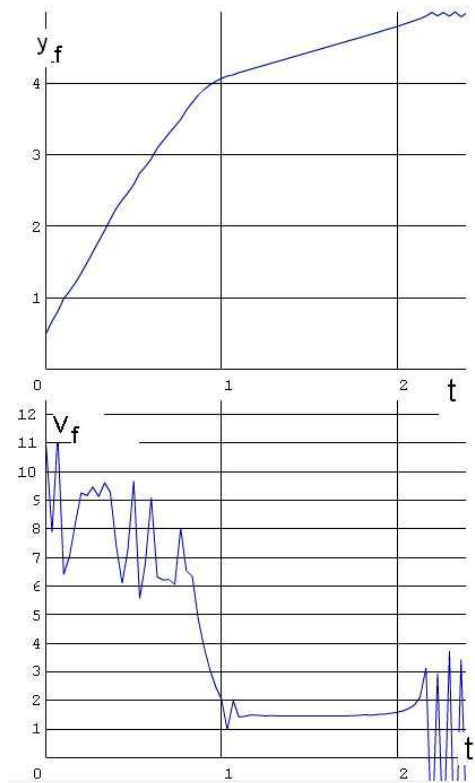} 
\end{center}
\caption{(Color online). The location $y_{f}$ and the velocity $v_{f}$ of
the front of vacancy distribution versus time. One can see two different
regimes of vacancy propagation through the sample: pulse driven at time $t<1$
and free expansion due to the Coulomb repulsion between the charged
vacancies ($t>1$) ending by establishing a stable distribution due to
vacancy confinement between sample boundaries. In both cases expansion is
linear in time meaning that the diffusion driven expansion $\propto \protect%
\sqrt{t}$ is insignificant.}
\label{front}
\end{figure}

\begin{figure}[tbp]
\begin{center}
\includegraphics[angle=-90,width=0.80\textwidth]{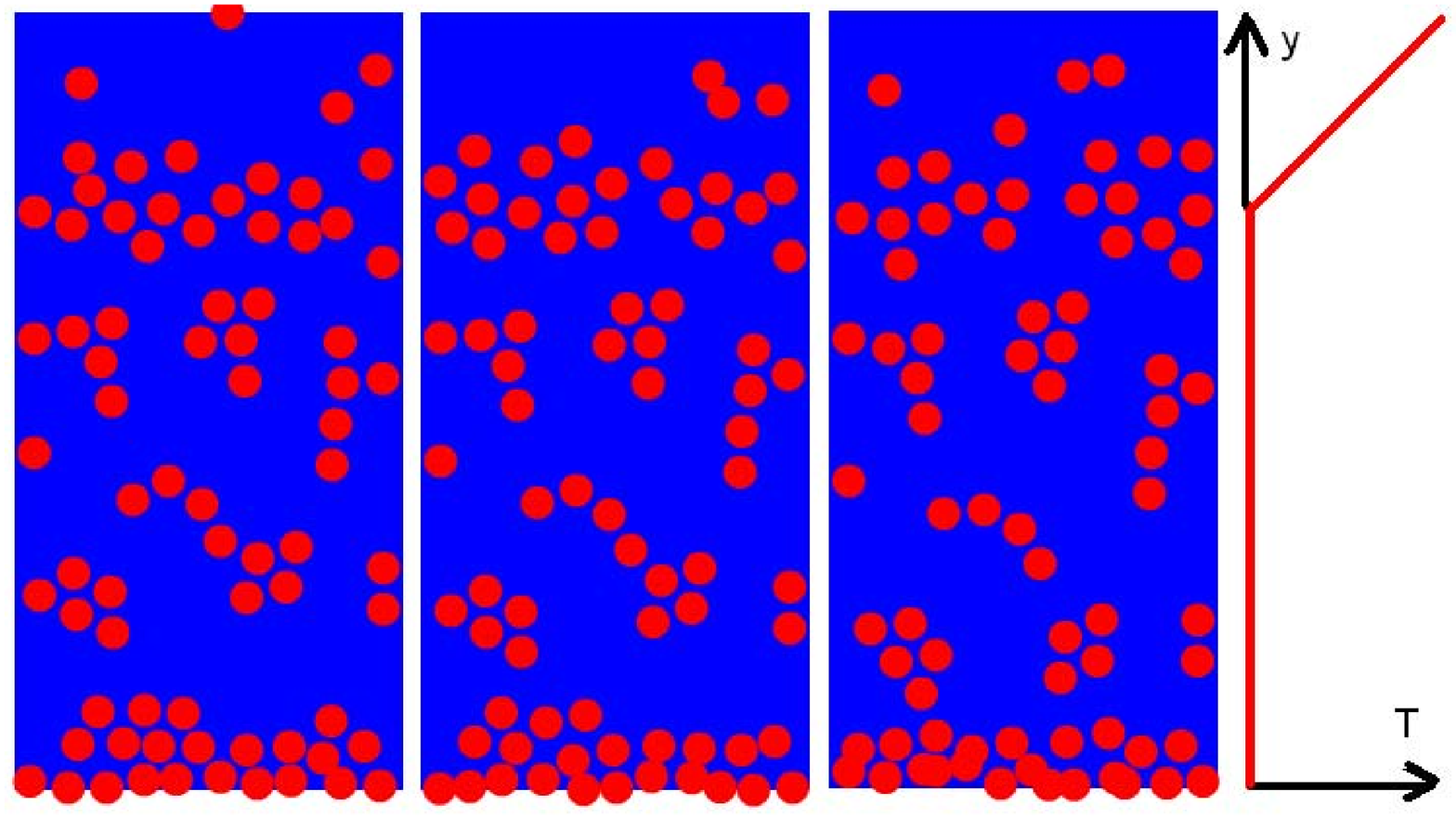}
\end{center}
\caption{(Color online) The same as in Fig.\protect\ref{fig:2Tx} for time ($%
t=$5; 5.5; 6), but now with a temperature gradient along the sample ($y-$%
axis) $T=T_{0}(5y-4h)\Theta (y-4h/5)/h$ and $T_{0}$ is five times
larger than the potential minimum depth. No stable configuration is
formed in the hot top part of the sample. The temperature profile
$T=T(y)$ is shown in the right sidebar. } \label{fig:3Ty}
\end{figure}
Comparison of our simulations with the averaged density description
indicates the following: (i) modeling vacancy dynamics by using macroscopic
distributions fails to describe filamentary cluster formation, (ii) there
are plenty of metastable vacancy distributions and, thus, one can expect
formation of slightly different percolation paths even for a series of
experiments on the same sample, (iii) MD simulations are ideal to describe
the statistics of such percolation path formation. Further deviation from
the standard diffusion approach can be seen when monitoring motion of
vacancy distribution front $y_{f}=\max_{i}(y_{i})$ (see Fig.~\ref{front}),
where the maximum is taken with respect to all the vacancies. One can see
two different expansion regimes for the front: a pulse driven regime and
another one driven by the Coulomb repulsion between the vacancies. Note that
the velocity of expansion is always constant indicating that diffusion
driven expansion is negligible in our system.

The temperature \emph{gradients} can change the vacancy distribution
patterns significantly. As examples, we have considered variations of the
temperature parallel to the electrodes, $T(x)$, Fig.~\ref{fig:2Tx}, and
across $T(y)$, Fig.~\ref{fig:3Ty}, our toy memristor. These gradients can be
due to, e.g., spatial variations of cooling or applying the heat sinks to
the sample. We use a square temperature profile along the film $%
T(x)=T_{0}\Theta (d/4-|x|)$ in Fig.~\ref{fig:2Tx}, where $x=0$ corresponds
to the middle of the sample, $d$ is the width of this region, and $\Theta $
is the step function. This mimics possible temperature oscillations along
the sample footprint. In the present study, we assume a temperature
independent mobility [i.e. a constant $\eta $ in Eq.~(\ref{lan})], which
would formally correspond to quantum tunneling rather than thermally
activated hopping. One can see that even in this regime the vacancies tend
to escape from the hot regions, resulting in a more pronounced formation of
percolation vacancy channels in cold regions of the simulated sample.

A somewhat similar situation occurs if we assume that the top of the sample
is hot with respect to the rest of the sample, i.e., $T(y)=T_{0}(y-y_{0})%
\Theta (y-y_{0})/(h-y_{0})$. In other words, we assume that the sample is
cold at the bottom ($T=0$ for $y<y_{0}$) and the temperature linearly
increases in the region $y_{0}<y<h$, where $h$ is the sample thickness. The
simulation results for such a distribution of temperature are shown in Fig.~%
\ref{fig:3Ty}. As one can see, vacancies cannot form stable configurations
in the hot region near the top of the sample: only limited number of
vacancies can enter the hot region, these vacancies quickly wander through
this region and return back into the cold part of the sample. In contrast,
the vacancy distribution is quite stable in the bottom \emph{cold} region of
the sample, as expected.

One can clearly see in Figs.~\ref{fig:2Tx} and \ref{fig:3Ty} that even a
moderate temperature/diffusion gradient drastically affects the formation of
percolation paths. Comparing with the homogeneous temperature regime, Fig.~%
\ref{fig:1puls}, the vacancies are escaping hot regions of the sample and
aggregating into the chain formations in the cold regions. This suggests a
strong competition between the local heating and the attractive part of the
vacancy-vacancy interaction resulting in disintegration/creation of the
conducting vacancy channels. Note that we do not consider interactions of
vacancies with the crystal matrix, which can be incorporated by adding a
crystal potential $U_{\mathrm{cr}}(x,y)$ in our Langevin equations. If
potential barriers of $U_{\mathrm{cr}}(x,y)$ are high enough and the motion
of vacancies occurs only via thermal hopping between nearest potential
minima of $U_{\mathrm{cr}}(x,y)$, then an exponential temperature dependence
of the effective mobility of vacancies can be observed. Thus, the
competition between the local heating and vacancy interactions should be
much stronger in the thermally activated (hopping) regime, when the
effective mobility (proportional to $\eta $) and the effective diffusion
coefficient exponentially depend on temperature \cite{md3}.

In conclusion, we have demonstrated that temperature gradients across
memristive devices, producing strong variations of the diffusion rate,
affect the vacancy patterns in memristors quite dramatically. Our
simulations indicate that variations of temperature across the memristor
favor the formation of short-circuiting or shunting vacancy channels. At the
same time, considerable temperature gradients in the growth direction, in
contrast, produce vacancy-poor regions where vacancy shunts are not formed
and electron percolation paths are blocked. Further studies of these effects
may have profound impact on present understanding of physics of memristive
devices.

\end{document}